\documentclass[letter]{IEEEtran}
\usepackage{graphicx,color}
\usepackage{cite}
\usepackage{setspace} 
\usepackage{amsmath,amsthm,amssymb}
\usepackage{multirow}
\usepackage{hhline}
\usepackage{epsfig}
\usepackage{epstopdf}
\usepackage{verbatim}
\usepackage{algorithm}
\usepackage{algpseudocode}
\usepackage{cases}
\usepackage{array}
\usepackage{mathptmx}
\usepackage[mathcal]{euscript}

\newtheorem{lemma}{Lemma}

\newtheorem{theorem}{Theorem}
\newtheorem{proposition}{Proposition}
\newtheorem{remark}{Remark}

\allowdisplaybreaks

\usepackage{forloop}
\newcounter{ct}

 \makeatletter 
 \def\@eqnnum{{\normalsize \normalcolor (\theequation)}} 
  \makeatother
\begin{document}

\title{OFDMA-based DF Secure Cooperative Communication with Untrusted Users\thanks{Manuscript received December 12, 2015; revised January 19, 2016; accepted January 21, 2016. This work has been supported by the Department of Science and Technology (DST) under Grant SB/S3/EECE/0248/2014. The associate editor coordinating the review of this paper and approving it for publication was K. Tourki.}}

\author{Ravikant Saini, {\em{\small Student Member, IEEE}}, Deepak Mishra, {\em{\small Student Member, IEEE}}, and Swades De, {\em{\small Senior Member, IEEE}}\thanks{R. Saini, D. Mishra, and S. De are with the Department of Electrical Engineering and Bharti School of Telecom,  Indian Institute of Technology Delhi, New Delhi, India (e-mail: ravikant.saini@dbst.iitd.ac.in; deepak.mishra@ee.iitd.ac.in; swadesd@ee.iitd.ac.in).}
\thanks{Digital Object Identifier xxxxxxxxxxxxxxxxx}
}
\maketitle

\begin{abstract}
In this letter we consider resource allocation for OFDMA-based secure cooperative communication by employing a trusted Decode and Forward (DF) relay among the untrusted users. We formulate two optimization problems, namely, (i) sum rate maximization subject to individual power constraints on source and relay, and (ii) sum power minimization subject to a fairness constraint in terms of per-user minimum support secure rate requirement. The optimization problems are solved utilizing the optimality of KKT conditions for pseudolinear functions.   
\end{abstract}

\section{Introduction}
\label{sec_introduction}

Relaying along with OFDMA is being considered as a promising technology for providing high data rate connectivity anywhere, anytime\cite{Salem_TCS_2010}. 
Physical layer security aspects in relay-assisted communication has recently gathered considerable attention in the research community \cite{amitav_TCST_2014}. Based on the relaying strategy, e.g., amplify-and-forward (AF) or decode-and-forward (DF), resource allocation problems are formulated differently and are thus investigated separately. 
Broadly, there exist two kinds of wire tapping scenarios: single eavesdropper with trusted users \cite{Jindal_CL_2015, Jeong_TSP_2011, Derrick_TWC_2011} and untrusted users \cite{RSAINI_TIFS_2016}. The study in \cite{Jindal_CL_2015} considered subcarrier and power allocation problems in an AF relay-assisted OFDM system with single eavesdropper. Assuming availability of direct path, \cite{Jeong_TSP_2011} considered sum rate maximization problem under total system power constraint in DF relay-assisted secure cooperative communication (DFSCC) for a single source-destination pair with a single eavesdropper. Multiuser resource allocation problem in OFDMA-based DFSCC with single eavesdropper was solved in \cite{Derrick_TWC_2011}. Recently, resource allocation problems for improving secure capacity and system fairness in OFDMA system with untrusted users and single friendly jammer have been considered in \cite{RSAINI_TIFS_2016}. \emph{To the best of our knowledge, OFDMA-based DFSCC with multiple untrusted users has not yet been considered in the literature.}

We consider two resource allocation problems. First, sum secure rate maximization is studied subject to individual power constraints on source and relay, due to their geographically apart locations. Second, sum power minimization is solved subject to the fairness constraint in terms of per-user minimum support secure rate. The key contributions are as follows: (i) We derive secure rate positivity constraints for each subcarrier, which includes the optimal subcarrier allocation policy. (ii) We prove that the two problems described above belong to the class of generalized convex problems which can be solved optimally. (iii) We show that the optimal secure rate for a user is achieved when rates of source-relay and relay-user links over a subcarrier are equal. (iv) We also present analytical and graphical interpretation of the derived optimal solutions. 

\section{System Model}
\label{sec_system_model}
We consider the downlink of an OFDMA-based cooperative communication system with a trusted DF relay controlled by a base station (hereafter referred as source $\mathcal{S}$). The users have mutual untrust and request secure communication from $\mathcal{S}$. The subcarriers on  $\mathcal{S}$-to-$\mathcal{R}$ and $\mathcal{R}$-to-$m$th user ($\mathcal{U}_m$) links are assumed to follow quasi-static Rayleigh fading. Availability of perfect CSI for each link is assumed. All nodes are equipped with single antenna, and $\mathcal{R}$ operates in half-duplex mode\cite{Jindal_CL_2015, Jeong_TSP_2011}. There is no direct connectivity between $\mathcal{S}$ and $\mathcal{U}_m$ \cite{Derrick_TWC_2011}. 

DFSCC with trusted $\mathcal{R}$ and $M$ untrusted users is a multiple eavesdropper scenario, where for each user there exist $M-1$ eavesdroppers, and the strongest of them is considered as the equivalent eavesdropper. Over a subcarrier $n$, the secure rate $R_{s_{n}}^m$ of $\mathcal{U}_m$ is defined as the non-negative difference of the rate $R_{n}^m$ of $\mathcal{U}_m$ and the rate $R_{n}^e$ of the equivalent  eavesdropper $\mathcal{U}_e$ \cite{RSAINI_TIFS_2016}. Mathematically, the secure rate is expressed as:
\begin{align}\label{rate_definition_1}
R_{s_{n}}^m = \left \{ R_{n}^m - \max\limits_{o \in \{1,2,\cdots M\} \setminus m} R_{n}^o \right \}^+ = \left\{ R_{n}^m - R_{n}^e \right \}^+
\end{align}
where $x^+ \hspace{-0.25mm}=\hspace{-0.25mm} \max\{0,x\}$. 
In half-duplex DF cooperative communication, $R_{n}^m = \frac{1}{2} \min \left \{ R_{n}^{sr},R_{n}^{rm} \right \}$, where $R_{n}^{sr}$ and $R_{n}^{rm}$ respectively denote the rates of $\mathcal{S}$-to-$\mathcal{R}$ and $\mathcal{R}$-to-$\mathcal{U}_m$ links over subcarrier $n$.
Using this, (\ref{rate_definition_1}) can be simplified as \cite{Jeong_TSP_2011}:
\begin{equation}\label{rate_definition_4}
R_{s_{n}}^m = \left(1/2\right) \left\{ \min{ (R_{n}^{sr}, R_{n}^{rm} ) } - R_{n}^{re} \right\}^+.
\end{equation}
Next, we discuss the sum secure rate maximization problem.

\section{Sum Secure Rate Maximization in DFSCC}
\label{sec_sum_rate_imp}
Denoting $P_{n}^s$ and $P_{n}^r$ respectively as powers of $\mathcal{S}$ and $\mathcal{R}$ over subcarrier $n$, the optimization problem can be stated as:
\begin{align}\label{opt_prob_rate_max}
& \mathcal{P}0: & & \underset{\pi_{n}^m, P_{n}^s, P_{n}^r} {\text{maximize}} 
\left[ R_s\left(\pi_{n}^m, P_{n}^s, P_{n}^r\right) = \sum_{m=1}^M \sum_{n=1}^N   \pi_{n}^m  R_{s_{n}}^m \right] \nonumber \\
& \text{s.t.} 
& & C1: \sum_{m=1}^M \pi_{n}^m \leq 1 \text{ } \forall n, \quad C2: \pi_{n}^m \in \{0,1\} \text{ } \forall m, n,\nonumber \\
& 
& & C3: \sum_{n=1}^N P_{n}^s \leq P_S, \quad \quad C4: \sum_{n=1}^N P_{n}^r \leq P_R, \nonumber \\
&
& & C5: P_{n}^s\ge 0, P_{n}^r\ge 0 \text{ } \forall n
\end{align}
where $\pi_{n}^m$ is a subcarrier allocation variable, indicating whether subcarrier $n$ is allocated to $\mathcal{U}_m$ or not. Constraints $C1$ and $C2$ ensure that a subcarrier is allocated to only one user. Power budgets $P_S$ and $P_R$ at $\mathcal{S}$ and $\mathcal{R}$ are respectively incorporated in $C3$ and $C4$. $C5$ includes positivity constraints. For each subcarrier, there are two real variables $P_{n}^s$, $P_{n}^r$, and one binary variable $\pi_{n}^m$. Because of $\log$ and $\max$ functions in objective, $\mathcal{P}0$ is a mixed integer non-linear programming problem, which is NP hard. To solve $\mathcal{P}0$, first we determine subcarrier allocation and then we complete power allocation. 

\subsection{Subcarrier Allocation}\label{subsec_subcarr_alloc}
The feasibility of achieving positive secure rate by $\mathcal{U}_m$ over a subcarrier $n$ is described by the following proposition.
\begin{proposition}\label{proposition_rate_positivity}
In DFSCC with untrusted users, positive secure rate over a subcarrier $n$ can be obtained if and only if (i) the subcarrier is allocated to the best gain user, and (ii) $\mathcal{R}$-to-$\mathcal{U}_e$ link of the eavesdropper $\mathcal{U}_e$ is the bottleneck link compared to the $\mathcal{S}$-to-$\mathcal{R}$ link over that subcarrier.
\end{proposition}
\begin{IEEEproof}
$R_{s_{n}}^m$ in (\ref{rate_definition_4}) can be restated as:
\begin{align}\label{simplified_sec_rate_def}
R_{s_{n}}^m = \frac{1}{2}
\begin{cases}
R_{n}^{sr} - R_{n}^{re} & \text{if $R_{n}^{re}<R_{n}^{sr}<R_{n}^{rm}$ }\\
R_{n}^{rm} - R_{n}^{re} & \text{if $R_{n}^{re}<R_{n}^{rm}<R_{n}^{sr}$ }\\
0 & \text{otherwise}.
\end{cases} 
\end{align}
From (\ref{simplified_sec_rate_def}) we note that, conditions for positive secure rate are: (a) $R_{n}^{re}<R_{n}^{rm}$ and (b) $R_{n}^{re}<R_{n}^{sr}$.  Let $\gamma_{n}^{sr}$, $\gamma_{n}^{rm}$, and $\gamma_{n}^{re}$ respectively  denote the channel gains of $\mathcal{S}$-to-$\mathcal{R}$,  $\mathcal{R}$-to-$\mathcal{U}_m$, and  $\mathcal{R}$-to-$\mathcal{U}_e$ links over subcarrier $n$. 
The rates $R_{n}^{sr}$, $R_{n}^{rm}$, and $R_{n}^{re}$ are given by 
$\log_2 \left( 1+ P_{n}^s\gamma_{n}^{sr}/{\sigma^2}\right)$, 
$\log_2 \left( 1+ P_{n}^r\gamma_{n}^{rm}/{\sigma^2}\right)$, and
$\log_2 \left( 1+ P_{n}^r\gamma_{n}^{re}/{\sigma^2}\right)$, respectively. 
Here $\sigma^2$ is the additive white Gaussian noise (AWGN) variance. Condition (a) $R_{n}^{re}<R_{n}^{rm}$, simplified as $\gamma_{n}^{re} < \gamma_{n}^{rm}$, indicates the optimal subcarrier allocation policy which can be stated as:
\begin{align}\label{subcarrier_alloc_relay}
\pi_{n}^m=
\begin{cases}
1 & \text{if $\gamma_{n}^{rm} > \gamma_{n}^{re}\triangleq\max\limits_{o \in \{1,2,\cdots M\} \setminus m} \gamma_{n}^{ro}$}\\
0 & \text{otherwise.}
\end{cases} 
\end{align}
Condition (b) $R_{n}^{re}<R_{n}^{sr}$, simplified as ${P_{n}^r \gamma_{n}^{re}}/\sigma^2<{P_{n}^s \gamma_{n}^{sr}}/\sigma^2$, should be incorporated as a power optimization constraint. \end{IEEEproof} 

Following the observations  $R_{n}^{re}\hspace{-0.5mm}<\hspace{-0.5mm}R_{n}^{rm}$ and $R_{n}^{re}\hspace{-0.5mm}<\hspace{-0.5mm}R_{n}^{sr}$ in Proposition \ref{proposition_rate_positivity}, $R_{s_{n}}^m$  can be rewritten without $\max$ operator as:
\begin{align}\label{simplified_objective}
R_{s_{n}}^m = \frac{1}{2} \left[ \log_2 \left( \frac{1+ \min{ \left( \frac{P_{n}^s\gamma_{n}^{sr}}{\sigma^2}, \frac{P_{n}^r \gamma_{n}^{rm}}{\sigma^2} \right)}} {1+\frac{P_{n}^r \gamma_{n}^{re}}{\sigma^2}} \right) \right].
\end{align}

\subsection{Power Allocation}\label{subsec_power_alloc}
Ensuring $R_{n}^{re}\hspace{-0.5mm}<\hspace{-0.5mm}R_{n}^{rm}$ through optimal subcarrier allocation \eqref{subcarrier_alloc_relay} and enforcing  $R_{n}^{re}\hspace{-0.5mm}<\hspace{-0.5mm}R_{n}^{sr}$ as a constraint, equivalent power allocation problem for $\mathcal{P}0$ using \eqref{simplified_objective} can be formulated as:
\begin{align}\label{opt_prob_rate_max_simplified_obj}
&\mathcal{P}1:\hspace{-3mm}&&\underset{P_{n}^s, P_{n}^r, t_n} {\text{maximize}} \left[ \widehat{R_{s}}(t_n, P_n^r) \triangleq  \sum_{n=1}^N \frac{1}{2} \left \{ \log_2 \left( \frac{1+ t_n} {1+ \frac{P_{n}^r \gamma_{n}^{re}}{\sigma^2}} \right) \right \}  \right] \nonumber \\
& \text{s.t.}  
& & C1: t_n \leq \frac{P_{n}^s\gamma_{n}^{sr}}{\sigma^2} \text{ } \forall n,\quad   C2: t_n \leq \frac{P_{n}^r\gamma_{n}^{rm}}{\sigma^2} \text{ } \forall n, \nonumber \\
&
& & C3, C4, C5, \text{as in } \eqref{opt_prob_rate_max},
\quad C6: \frac{P_{n}^r \gamma_{n}^{re}}{\sigma^2} \le \frac{P_{n}^s \gamma_{n}^{sr}}{\sigma^2}\text{ } \forall n .  
\end{align} 
Constraints $C1-C2$ come from the definition of  $\min\left\lbrace\cdot\right\rbrace$, $C6$ comes from  secure rate positivity requirements given by Proposition \ref{proposition_rate_positivity}. Due to non-concave objective function $\widehat{R_{s}}$, $\mathcal{P}1$ is non-convex. However, via the following lemma, we show that $\mathcal{P}1$ belongs to the class of generalized convex problems. 
\begin{lemma}\label{PL:lemma}
The objective function of $\mathcal{P}1$ is pseudolinear on the feasible region defined by the constraints, and the solution obtained from the KKT conditions is the global optimal.
\end{lemma}
\begin{IEEEproof}
The objective function $\widehat{R_{s}}\left(t_n, P_{n}^r\right)$ of $\mathcal{P}1$ is a pseudolinear function~\cite{komlosi1993first} of $t_{n}$ and $P_{n}^r$, with $\frac{\partial \widehat{R_{s}}}{\partial t_{n}} = \frac{1}{2(1+t_n)} \triangleq a_n$ and $\frac{\partial \widehat{R_{s}}}{\partial P_{n}^r} =  \frac{-\gamma_{n}^{re}}{2(\sigma^2+P_{n}^r\gamma_{n}^{re})} \triangleq b_n$, because $\frac{\partial \widehat{R_{s}}}{\partial t_{n}}, \frac{\partial \widehat{R_{s}}}{\partial P_{n}^r}\neq 0$ in the entire feasible region defined by the linear constraints $C1-C6$.
Moreover, determinant of the bordered Hessian is given as:
$Det(B_{H})=
\begin{vmatrix}
0 & \frac{\partial \widehat{R_{s}}}{\partial t_n} & \frac{\partial \widehat{R_{s}}}{\partial P_n^r}\\
\frac{\partial \widehat{R_{s}}}{\partial t_n} & \frac{\partial^2 \widehat{R_{s}}}{\partial t_n^2} & \frac{\partial^2 \widehat{R_{s}}}{\partial {t_n}\partial P_n^r} \\
\frac{\partial \widehat{R_{s}}}{\partial P_n^r} & \frac{\partial^2 \widehat{R_{s}}}{\partial P_n^r \partial {t_n}} & \frac{\partial^2 \widehat{R_{s}}}{\partial {P_n^r}^2} \end{vmatrix}
= 
\begin{vmatrix}
0 & a_n & b_n\\
a_n & -a_n^2 & 0 \\
b_n & 0 & b_n^2 \end{vmatrix}
 = 0.$\\
Following this and \cite[Theorem 4.3.8]{Bazaraa}, along with the knowledge that constraints are linear and differentiable, it can be inferred that the KKT point is the global optimal solution.
\end{IEEEproof}

The optimal power allocation ($P_{n}^s$, $P_{n}^r$) obtained after solving KKT conditions is characterized by following Theorem.

\begin{theorem}\label{theorem_one}
In DFSCC, maximum secure rate over a subcarrier is achieved when the following relationship holds
\begin{equation}
\label{max_capacity_condition}
P_{n}^s \gamma_{n}^{sr} = P_{n}^r \gamma_{n}^{rm}.
\end{equation}
\end{theorem}
\begin{IEEEproof}
Lagrangian $\mathcal{L}_1$ of the problem $\mathcal{P}1$ can be stated as: 
\begin{align}\label{lagrange_rate_maximization}
\mathcal{L}_1 = &\sum_{n=1}^N \frac{1}{2} \left \{ \log_2 \left( \frac{1+ t_n} {1+ \frac{P_{n}^r \gamma_{n}^{re}}{\sigma^2}} \right) \right \} - \sum_{n=1}^N x_n \left( t_n - \frac{P_{n}^s\gamma_{n}^{sr}}{\sigma^2} \right) \nonumber\\
 & - \sum_{n=1}^N y_n \left( t_n - \frac{P_{n}^r\gamma_{n}^{rm}}{\sigma^2} \right)
  - \sum_{n=1}^N z_n \left( \frac{P_{n}^r \gamma_{n}^{re}}{\sigma^2} - \frac{P_{n}^s \gamma_{n}^{sr}}{\sigma^2} \right) \nonumber\\  
&- \lambda \left( \sum_{n=1}^N P_{n}^s - P_S \right)  - \mu \left( \sum_{n=1}^N P_{n}^r - P_R \right).
\end{align}

Here, $x_n, y_n, z_n, \lambda$, and $\mu$ are Lagrange multipliers. Using (\ref{opt_prob_rate_max_simplified_obj}) and (\ref{lagrange_rate_maximization}), the KKT conditions for $\mathcal{P}1$ are given by
\begin{subequations}
\begin{equation}\label{lang_derivative_psn}
\frac{\partial \mathcal{L}_1}{\partial P_{n}^s} =  x_n \frac{\gamma_{n}^{sr}}{\sigma^2} + z_n \frac{\gamma_{n}^{sr}}{\sigma^2} -\lambda = 0
\end{equation}
\begin{equation}\label{lang_derivative_prn}
\frac{\partial \mathcal{L}_1}{\partial P_{n}^r} = \frac{-\gamma_{n}^{re}}{2\left( \sigma^2+P_{n}^r\gamma_{n}^{re} \right) }  + y_n \frac{\gamma_{n}^{rm}}{\sigma^2} - z_n \frac{\gamma_{n}^{re}}{\sigma^2} -\mu = 0
\end{equation} 
\begin{equation}\label{lang_derivative_zn}
\frac{\partial \mathcal{L}_1}{\partial t_{n}} = \frac{1}{2\left( 1+t_n \right)} - x_n -y_n = 0
\end{equation} 
\begin{equation}\label{comp_slackness_dummy_var}
x_n \left( t_n - \frac{P_{n}^s\gamma_{n}^{sr}}{\sigma^2} \right) = 0; \quad y_n \left( t_n - \frac{P_{n}^r\gamma_{n}^{rm}}{\sigma^2} \right) = 0
\end{equation}
\begin{equation}\label{comp_slackness_power_contraint}
z_n \left( {P_{n}^r \gamma_{n}^{re}} - {P_{n}^s \gamma_{n}^{sr}}\right) = 0
\end{equation}
\begin{equation}\label{comp_slackness_power}
\lambda \left( \sum_{n=1}^N P_{n}^s - P_S\right) = 0; \quad \mu \left(\sum_{n=1}^N P_{n}^r - P_R \right) = 0.
\end{equation}
\end{subequations}

\noindent Next we consider the following three cases.
\begin{align}\label{three_cases_for_t_n}
t_{n} = 
\begin{cases}
P_{n}^s\gamma_{n}^{sr}/\sigma^2  & \text{if $P_{n}^s\gamma_{n}^{sr} < P_{n}^r\gamma_{n}^{rm}$ }\\
P_{n}^r\gamma_{n}^{rm}/\sigma^2  & \text{if $P_{n}^s\gamma_{n}^{sr} > P_{n}^r\gamma_{n}^{rm}$ }\\
P_{n}^s\gamma_{n}^{sr}/\sigma^2 = P_{n}^r\gamma_{n}^{rm}/\sigma^2 & \text{otherwise.}
\end{cases} 
\end{align}

Case 1: 
$t_{n} = P_{n}^s\gamma_{n}^{sr}/ \sigma^2$; $x_n>0$ and $y_n=0$.
From (\ref{lang_derivative_prn}) we get $\frac{\gamma_{n}^{re}}{2\left( \sigma^2+P_{n}^r\gamma_{n}^{re} \right) } + \mu  + z_n \frac{\gamma_{n}^{re}}{\sigma^2} = 0$, which cannot be satisfied for any positive finite $P_{n}^r$ and $\sigma^2$. Thus, this case is infeasible.

Case 2: 
$t_{n} = P_{n}^r\gamma_{n}^{rm} / \sigma^2$; $x_n=0$ and $y_n>0$, it gives $\lambda = z_n \gamma_{n}^{sr}/\sigma^2$ and $y_n = 1/\left\{ 2(1+t_n)\right \}$.
Substituting in (\ref{lang_derivative_prn}),
\begin{align}\label{opt_power_solution_case_2}
\mu + \lambda \frac{\gamma_{n}^{re}}{\gamma_{n}^{sr}} + \frac{\gamma_{n}^{re}}{2 \left( \sigma^2+P_{n}^r\gamma_{n}^{re} \right) } - \frac{\gamma_{n}^{rm}}{2 \left( \sigma^2+P_{n}^r\gamma_{n}^{rm} \right) } = 0.
\end{align}
Since $P_{n}^r\gamma_{n}^{rm}/\sigma^2 < P_{n}^s\gamma_{n}^{sr}/\sigma^2$, and we know that $\gamma_{n}^{rm}>\gamma_{n}^{re}$. Thus, $P_{n}^r\gamma_{n}^{re}/\sigma^2 < P_{n}^r\gamma_{n}^{rm}/\sigma^2 < P_{n}^s\gamma_{n}^{sr}/\sigma^2$ which indicates that $z_n=0$ (c.f. (\ref{comp_slackness_power_contraint})).  Since $z_n=0$, so $\lambda=0$. Thus, (\ref{opt_power_solution_case_2}) results in a quadratic equation in $P_{n}^r$, having  following form
\begin{align}\label{quadratic_prn}
\left(P_{n}^r\right)^2 \gamma_{n}^{rm} \gamma_{n}^{re} + P_{n}^r \left(\gamma_{n}^{rm} +  \gamma_{n}^{re}\right)\sigma^2 + \sigma^4 - \Delta_n = 0 
\end{align}
where  $\Delta_n = \frac{\sigma^2}{2\mu}\left(\gamma_{n}^{rm} -  \gamma_{n}^{re}\right)$.
For a fixed $\mu$, the optimal $P_{n}^r$, obtained as the only positive real root of (\ref{quadratic_prn}) is given by
\begin{align}\label{optimal_solution_prn}
\hspace{-2mm}P_{n}^r\hspace{-0.5mm} =\hspace{-0.5mm}\frac{-\sigma^2(\gamma_{n}^{rm}\hspace{-0.5mm}+\hspace{-0.5mm}\gamma_{n}^{re})
\hspace{-0.5mm}+\hspace{-0.5mm} \sqrt{\hspace{-0.5mm}\sigma^4 \left(\gamma_{n}^{rm}\hspace{-0.5mm}-\hspace{-0.5mm}\gamma_{n}^{re} \right)^2 \hspace{-0.5mm}+\hspace{-0.5mm} 4 \gamma_{n}^{rm}\gamma_{n}^{re}\Delta_n } }{2\gamma_{n}^{rm}\gamma_{n}^{re}}\hspace{-0.5mm}.\hspace{-2mm}
\end{align}
For a known $P_{n}^r$, we set $P_{n}^s = P_{n}^r\frac{\gamma_{n}^{rm}}{\gamma_{n}^{sr}} + \delta$, where $\delta$ is a very small positive number, because allocating more $P_{n}^s$ does not provide a higher secure rate. The optimum $\mu$ can be found using subgradient method \cite{sboyd_subgradient}, such that $\sum_{n=1}^N P_{n}^r = P_R$. 

Case 3: 
$t_{n} = P_{n}^r\gamma_{n}^{rm}/\sigma^2 = P_{n}^s\gamma_{n}^{sr}/\sigma^2$; 
$x_n>0$ and $y_n>0$. Replacing $x_n = \lambda\frac{\sigma^2}{\gamma_{n}^{sr}} - z_n$ (from (\ref{lang_derivative_psn})), 
in (\ref{lang_derivative_zn}) we get
$y_n \hspace{-0.5mm}=\hspace{-0.5mm} \frac{1}{2\left( 1+t_n \right) } \hspace{-0.5mm}-\hspace{-0.5mm} \lambda\frac{\sigma^2}{\gamma_{n}^{sr}} \hspace{-0.5mm}+\hspace{-0.5mm} z_n$.
On substituting $t_{n}$ and $y_n$ in (\ref{lang_derivative_prn}),
\begin{align}\label{opt_power_solution_case_3}
& \frac{\gamma_{n}^{re}}{2\left( \sigma^2+P_{n}^r\gamma_{n}^{re} \right) }- \frac{\gamma_{n}^{rm}}{2 \left( \sigma^2+P_{n}^r\gamma_{n}^{rm} \right) } + \mu +  \lambda \left( \frac{\gamma_{n}^{rm}} {\gamma_{n}^{sr}} \right) \nonumber \\ 
& \qquad \qquad = z_n \left(\gamma_{n}^{rm}-\gamma_{n}^{re} \right)/ \sigma^2.
\end{align}

The complimentary slackness condition  
(\ref{comp_slackness_power_contraint}) can be simplified as 
$z_n \left(P_{n}^r \gamma_{n}^{re}/\sigma^2 - P_{n}^r \gamma_{n}^{rm}/\sigma^2 \right) = 0$, which indicates $z_n=0$ because $\gamma_{n}^{rm}>\gamma_{n}^{re}$.
After simplifications, (\ref{opt_power_solution_case_3}) results in a quadratic equation in $P_{n}^r$ similar to (\ref{quadratic_prn}) with 
$\Delta_n = \frac{\sigma^2 \gamma_{n}^{sr} \left(\gamma_{n}^{rm} -  \gamma_{n}^{re} \right) }{2 \left( \mu\gamma_{n}^{sr}+\lambda\gamma_{n}^{rm} \right) }$. 
For fixed $\lambda$ and $\mu$, the optimal $P_n^r$ is given by (\ref{optimal_solution_prn}).
The optimal $\lambda$ and $\mu$ are obtained using subgradient method \cite{sboyd_subgradient} such that $\sum_{n=1}^N P_{n}^r = P_R$ and $\sum_{n=1}^N P_{n}^s = P_S$. 
Since with $\delta\to 0$ case 2 is inherently contained in case 3, (\ref{max_capacity_condition}) provides an energy-efficient global optimal solution. 
\end{IEEEproof}

\subsection{Analytical and Graphical Interpretations}\label{subsec_analytic_interp}
Writing (\ref{opt_power_solution_case_3}) with $z_n=0$ in a compact form we get 
\begin{align}\label{analytical_insight}
\frac{\frac{\gamma_{n}^{rm}-\gamma_{n}^{re}}{\sigma^2}} {2\left(1 + \frac{P_{n}^r\gamma_{n}^{rm}}{\sigma^2} \right) \left(1 + \frac{P_{n}^r\gamma_{n}^{re}}{\sigma^2} \right)} = \mu + \lambda \left( \frac{\gamma_{n}^{rm}}{\gamma_{n}^{sr}} \right) \text{ } \forall n.
\end{align}
From \eqref{simplified_objective} and \eqref{max_capacity_condition}, it appears intuitive that $P_{n}^r$ should be allocated according to the relative gain $(\gamma_{n}^{rm}\hspace{-0.5mm}-\hspace{-0.5mm}\gamma_{n}^{re})$. However on closely observing \eqref{analytical_insight}, we note that $P_{n}^r$ depends not only on  relative gain, but also on absolute gains $\gamma_{n}^{sr}$, $\gamma_{n}^{rm}$, and $\gamma_{n}^{re}$.

\begin{figure}[!t]
\centering
\epsfig{file=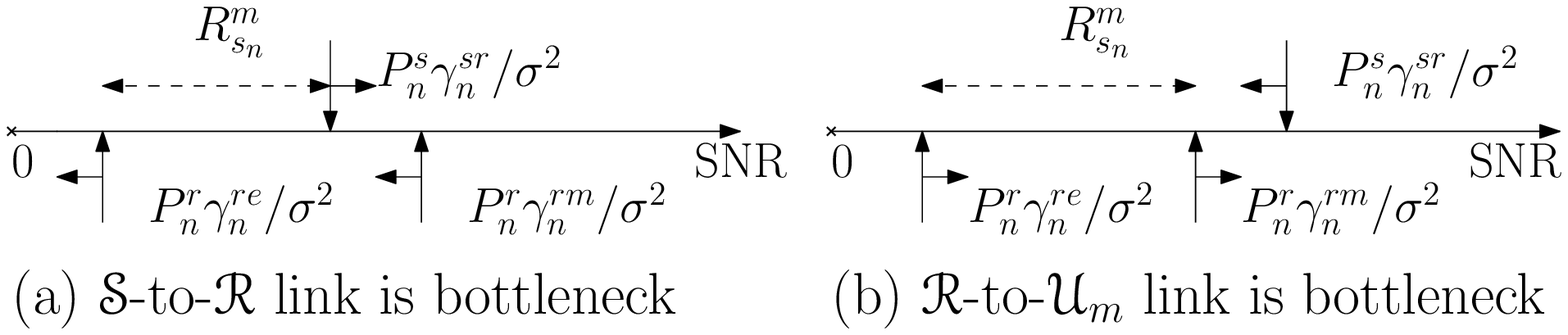,width=3.2in}
\vspace{-2mm}\caption{Graphical interpretation of Theorem \ref{theorem_one}.}
\label{fig:case3_explanation}
\end{figure}

Utilizing the secure rate definition (\ref{simplified_sec_rate_def}), the result \eqref{max_capacity_condition} obtained from Theorem \ref{theorem_one} can be explained graphically using Fig. \ref{fig:case3_explanation}. When $\mathcal{S}$-to-$\mathcal{R}$ link is the bottleneck as compared to $\mathcal{R}$-to-$\mathcal{U}_m$ link i.e., $P_{n}^s \gamma_{n}^{sr} < P_{n}^r \gamma_{n}^{rm}$ 
(case 1 in (\ref{three_cases_for_t_n}), and case (a) in Fig. \ref{fig:case3_explanation}), the secure rate is given as $R_{s_n}^m = \log_2\left(\frac{\sigma^2+P_{n}^s \gamma_{n}^{sr}}{\sigma^2+P_{n}^r \gamma_{n}^{re}}\right).$
This case is infeasible because $R_{s_n}^m$ can be improved by either increasing $P_{n}^s$ or reducing $P_{n}^r$. If there is enough $P_S$ budget, then $P_{n}^s$ could be increased till (\ref{max_capacity_condition}) gets satisfied, beyond which $\mathcal{R}$-to-$\mathcal{U}_m$ link becomes the bottleneck (considered separately as case (b) in Fig. \ref{fig:case3_explanation}). However, if $P_S$ is limited, then $P_{n}^r$ should be reduced till (\ref{max_capacity_condition}) gets satisfied. With further lowered $P_{n}^r$, $\mathcal{R}$-to-$\mathcal{U}_m$ link becomes the bottleneck. Thus, at  KKT point $P_{n}^s \gamma_{n}^{sr}$ cannot be less than $P_{n}^r \gamma_{n}^{rm}$, and hence case 1 is infeasible.

When $\mathcal{R}$-to-$\mathcal{U}_m$ link is bottleneck compared to $\mathcal{S}$-to-$\mathcal{R}$ link
i.e., $P_{n}^r \gamma_{n}^{rm} < P_{n}^s \gamma_{n}^{sr} $  (case 2 in (\ref{three_cases_for_t_n}), and case (b) in Fig. \ref{fig:case3_explanation}), $R_{s_n}^m = \log_2\left(\frac{\sigma^2+P_{n}^r \gamma_{n}^{rm}}{\sigma^2+P_{n}^r \gamma_{n}^{re}}\right)$,
which is an increasing function of $P_{n}^r$. In order to improve $R_{s_n}^m$, $P_{n}^r$ can be increased till (\ref{max_capacity_condition}) gets satisfied, beyond which the $\mathcal{S}$-to-$\mathcal{R}$ link becomes  bottleneck. If $P_R$ is limited then just enough $P_{n}^s$ should be utilized so as to satisfy (\ref{max_capacity_condition}). Higher $P_{n}^s$, though feasible, does not improve secure rate, as $R_{s_n}^m$ is independent of $P_{n}^s$. Thus, case 2 can have multiple solutions but with the same optimal value.
\begin{remark}\label{rem:rem1}
Using (\ref{max_capacity_condition}) in (\ref{simplified_objective}), at KKT point $R_{s_{n}}^m$ is concave increasing in $P_n^r$, and is  bounded by $\left(1/2\right)\log_2 \left( {\gamma_{n}^{rm}}/{ \gamma_{n}^{re}} \right)$.
\end{remark}

\section{Sum Power Minimization in DFSCC}
\label{sec_power_min_sum_rate}
For the sum power minimization problem, considering sum secure rate constraint over OFDMA system is not significant from fairness point of view. Actually it would utilize power resources over those subcarriers where higher secure rate can be achieved, which may lead to scarcity of power resources and eventually very small secure rate for some users. For handling this issue, we consider a minimum support secure rate requirement $R_{ssr}$  for each user. This results in a more complicated problem with $M$ rate constraints, instead of one.

Following Proposition \ref{proposition_rate_positivity}, first, subcarrier allocation is done based on (\ref{subcarrier_alloc_relay}). Since the maximum secure rate achievable over a subcarrier is bounded (cf. Remark~\ref{rem:rem1}), before optimal power allocation, it is checked whether $R_{ssr}$ for each user can be achieved. If it can be achieved, that user is selected for power allocation, otherwise not. Considering $U^a$ as the set of selected users for resource allocation and $N_m$ as the set of subcarriers of $\mathcal{U}_m$, the sum power minimization problem is given by:
\begin{align}\label{opt_prob_power_min}
&\hspace{-1mm}\mathcal{Q}0 : & & \underset{P_{k}^s, P_{k}^r} {\text{minimize}} \sum_{\mathcal{U}_m \in U^a} \sum_{k \in N_m} \left( P_{k}^s + P_{k}^r \right) \nonumber \\
&\hspace{-1mm}\text{s.t.}  
& & \hspace{-5mm}C1\hspace{-1mm}:\hspace{-1mm}\sum_{k \in N_m} R_{s_{k}}^m \geq R_{ssr} \text{ }\forall\hspace{0.5mm}\mathcal{U}_m\hspace{-0.5mm}\in\hspace{-0.5mm} U^a, \hspace{1mm} C2\hspace{-0.5mm}: P_{k}^s, P_{k}^r\ge 0. 
\end{align} 
Since all subcarriers are independent, per-user rate constraints can be handled in parallel. Thus, the optimization problem can be decomposed at user level and solved in parallel. The individual user level problem for each  $\mathcal{U}_m\in U^a$ is stated as:

\begin{align}\label{opt_prob_power_min_user}
&\mathcal{Q}1 : & & \underset{ P_{k}^r, P_{k}^s} {\text{minimize}} \sum_{k \in N_m} \left(P_{k}^s + P_{k}^r  \right)  \nonumber \\
& \text{s.t.} 
& & \hspace{-5mm}C1: \sum_{k \in N_m} \frac{1}{2} \log_2 \left( \frac{1+t_k}{1+\frac{P_{k}^r\gamma_{k}^{re}}{\sigma^2}} \right) \geq R_{ssr}, \nonumber \\
&
& & \hspace{-5mm}C2: t_k \leq \frac{P_{k}^s\gamma_{k}^{sr}}{\sigma^2} \text{ } \forall k, \quad C3: t_k \leq \frac{P_{k}^r\gamma_{k}^{rm}}{\sigma^2} \text{ } \forall k, \nonumber \\
&
& & \hspace{-5mm}C4: {P_{k}^r \gamma_{k}^{re}}/{\sigma^2} \le {P_{k}^s \gamma_{k}^{sr}}/{\sigma^2} \text{ } \forall k,  C5: P_{k}^s\ge 0, P_{k}^r\ge 0\text{ } \forall k.
\end{align}
The objective function of $\mathcal{Q}1$ is linear, $C1$ is pseudolinear (Lemma~\ref{PL:lemma}), and $C2-C5$ are linear. 
So, the KKT point gives the optimal solution \cite{Bazaraa}.
The Lagrangian $\mathcal{L}_2$ of $\mathcal{Q}1$ with $x_k, y_k, z_k$, and  $\lambda$ as the Lagrange multipliers is given by:
\begin{align}
&\mathcal{L}_2 =  \sum_{k \in N_m} \left(P_{k}^s + P_{k}^r\right) + \sum_{k \in N_m} x_k \left( t_k - \frac{P_{k}^s\gamma_{k}^{sr}}{\sigma^2} \right) \nonumber\\
 & + \sum_{k \in N_m} y_k \left( t_k - \frac{P_{k}^r\gamma_{k}^{rm}}{\sigma^2} \right)
  + \sum_{k \in N_m} z_k \left( \frac{P_{k}^r \gamma_{k}^{re}}{\sigma^2} - \frac{P_{k}^s \gamma_{k}^{sr}}{\sigma^2} \right) \nonumber\\  
& - \lambda \left[ \sum_{k \in N_m} \frac{1}{2} \left \{ \log_2 \left( \frac{1+ t_k} {1+ \frac{P_{k}^r \gamma_{k}^{re}}{\sigma^2}} \right) \right \} - R_{ssr} \right].
\end{align}
The stationarity KKT conditions for $\mathcal{Q}1$ are given by:
\begin{subequations}
\begin{equation}\label{lang_derivative_psn_1}
\frac{\partial \mathcal{L}_2}{\partial P_{k}^s} =  1 - x_k \frac{\gamma_{k}^{sr}}{\sigma^2} - z_k \frac{\gamma_{k}^{sr}}{\sigma^2}  = 0
\end{equation} 
\begin{equation}\label{lang_derivative_prn_1}
\frac{\partial \mathcal{L}_2}{\partial P_{k}^r} = 1 + \frac{\lambda \gamma_{k}^{re}}{2\left( \sigma^2+P_{k}^r\gamma_{k}^{re} \right) }  - y_k \frac{\gamma_{k}^{rm}}{\sigma^2} + z_k \frac{\gamma_{k}^{re}}{\sigma^2} = 0
\end{equation} 
\begin{equation}\label{lang_derivative_zn_1}
\frac{\partial \mathcal{L}_2}{\partial t_{k}} =  x_k + y_k - \frac{\lambda}{2 \left( 1+t_k \right) }  = 0.
\end{equation}
\end{subequations}
In addition, there are four complimentary slackness conditions, three are similar to  (\ref{comp_slackness_dummy_var})-(\ref{comp_slackness_power_contraint}), and fourth is given as:
\begin{align}\label{comp_slackness_rate}
\lambda \left[ \sum_{k \in N_m} \frac{1}{2} \left \{ \log_2 \left( \frac{1+ t_k} {1+ \frac{P_{k}^r \gamma_{k}^{re}}{\sigma^2}} \right) \right \} - R_{ssr} \right] = 0. 
\end{align}

Here also there exist three cases similar to (\ref{three_cases_for_t_n}).  Considering the cases one by one, in case 1, $x_k>0$ but $y_k=0$. This case is infeasible because, if $y_k=0$, then (\ref{lang_derivative_prn_1}) cannot be satisfied. Considering case 2, $x_k=0$, $y_k>0$, and $z_k=0$ (by the same argument as explained in the proof of Theorem \ref{theorem_one}); thus (\ref{lang_derivative_psn_1}) cannot be satisfied, and hence this case is also infeasible. The only feasible case is case 3, in which, using $z_k=0$ and on simplifying (\ref{lang_derivative_psn_1})-(\ref{lang_derivative_zn_1}), we obtain a quadratic equation in $P_{k}^r$ similar to (\ref{quadratic_prn}) where $\Delta_n$ is replaced with 
$\Delta_k = \frac{\lambda \sigma^2 \gamma_{k}^{sr} \left( \gamma_{k}^{rm} -  \gamma_{k}^{re} \right) }{2 \left( \gamma_{k}^{sr}+\gamma_{k}^{rm} \right)}.$ 
The optimal $P_{k}^r$ is given by (\ref{optimal_solution_prn}) for  a fixed value of $\lambda$, and the optimal $\lambda$ is found using subgradient method \cite{sboyd_subgradient} such that $C1$ in $\mathcal{Q}1$ is satisfied with equality. 

\section{Numerical Results}
\label{sec_results}
Here we present numerical results of OFDMA-based DFSCC with $M=8$ users and $N=64$ subcarriers. $\mathcal{S}$ and $\mathcal{R}$ are assumed to be respectively located at $(0,0)$ and $(1,0)$. The users are uniformly distributed inside a unit square, centered at $(2,0)$. With $\sigma^2=1$, we consider quasi-static Rayleigh fading. Large scale fading is modeled using path loss exponent $=3$.

\begin{figure}[t!]
\centering
\epsfig{file=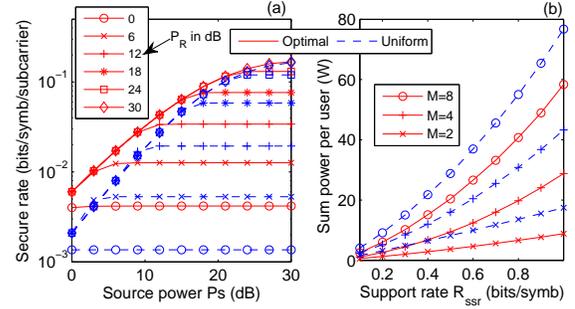,width=3.in}
\vspace{-2mm}\caption{\hspace{-1.5mm}(a)\hspace{0.5mm}Secure rate versus source power; (b)\hspace{0.5mm}Sum power versus support rate.}
\label{fig:sum_rate_max}
\end{figure}

Fig. \ref{fig:sum_rate_max}(a) presents the variation of optimal sum secure rate $R_s^*$ (or $\widehat{R_{s}^*}$) with source power budget $P_S$, for different relay power budgets $P_R$. At low $P_R$, $R_s^*$ is limited by $P_R$ itself and increasing $P_S$ does not improve $R_s^*$ significantly.  Interestingly, at high enough $P_R$, $R_s^*$ increases with diminishing returns before saturating at high $P_S$. This indicates the existence of an upper bound on $R_s^*$. {\em The monotonicity of $R_s^*$ corroborates pseudolinearity with respect to $P_n^s$ and $P_n^r$.} Fig. \ref{fig:sum_rate_max}(b) shows that sum power required per-user increases exponentially with $R_{ssr}$. Sum power required for a fixed $R_{ssr}$ increases with number of users, due to effectively lesser number of subcarriers per-user. The performance improvement achieved by the proposed solutions over  a benchmark scheme, namely, uniform power allocation, is also demonstrated in Figs. \ref{fig:sum_rate_max}(a) and \ref{fig:sum_rate_max}(b).
 
\section{Concluding Remarks} 	
\label{sec_conclusion}
To summarize, we have investigated resource allocation in OFDMA-based DFSCC with multiple untrusted users. Global optimal solutions for secure rate maximization and sum power minimization problems have been obtained by exploiting the concepts of generalized convexity and pseudolinearity. Numerical results have offered insight into the optimal power required for realizing an energy-efficient DFSCC system.



\begin{thebibliography}{1}
\providecommand{\url}[1]{#1}
\csname url@samestyle\endcsname
\providecommand{\newblock}{\relax}
\providecommand{\bibinfo}[2]{#2}
\providecommand{\BIBentrySTDinterwordspacing}{\spaceskip=0pt\relax}
\providecommand{\BIBentryALTinterwordstretchfactor}{4}
\providecommand{\BIBentryALTinterwordspacing}{\spaceskip=\fontdimen2\font plus
\BIBentryALTinterwordstretchfactor\fontdimen3\font minus
  \fontdimen4\font\relax}
\providecommand{\BIBforeignlanguage}[2]{{%
\expandafter\ifx\csname l@#1\endcsname\relax
\typeout{** WARNING: IEEEtran.bst: No hyphenation pattern has been}%
\typeout{** loaded for the language `#1'. Using the pattern for}%
\typeout{** the default language instead.}%
\else
\language=\csname l@#1\endcsname
\fi
#2}}
\providecommand{\BIBdecl}{\relax}
\BIBdecl

\bibitem{Salem_TCS_2010}
M.~Salem, A.~Adinoyi, M.~Rahman, H.~Yanikomeroglu, D.~Falconer, Y.-D. Kim,
  E.~Kim, and Y.-C. Cheong, ``An overview of radio resource management in
  relay-enhanced {OFDMA}-based networks,'' \emph{IEEE Commun. Surveys Tuts.},
  vol.~12, no.~3, pp. 422--438, Aug. 2010.

\bibitem{amitav_TCST_2014}
A.~Mukherjee, S.~Fakoorian, J.~Huang, and A.~Swindlehurst, ``Principles of
  physical layer security in multiuser wireless networks: A survey,''
  \emph{IEEE Commun. Surveys Tuts.}, vol.~16, no.~3, pp. 1550--1573, Aug. 2014.

\bibitem{Jindal_CL_2015}
A.~Jindal and R.~Bose, ``Resource allocation for secure multicarrier {AF} relay
  system under total power constraint,'' \emph{IEEE Commun. Lett.}, vol.~19,
  no.~2, pp. 231--234, Feb. 2015.

\bibitem{Jeong_TSP_2011}
C.~Jeong and I.-M. Kim, ``Optimal power allocation for secure multicarrier
  relay systems,'' \emph{IEEE Trans. Signal Process.}, vol.~59, no.~11, pp.
  5428--5442, Nov. 2011.

\bibitem{Derrick_TWC_2011}
D.~Ng, E.~Lo, and R.~Schober, ``Secure resource allocation and scheduling for
  {OFDMA} decode-and-forward relay networks,'' \emph{IEEE Trans. Wireless
  Commun.}, vol.~10, no.~10, pp. 3528--3540, Oct. 2011.

\bibitem{RSAINI_TIFS_2016}
R. Saini, A. Jindal, and S. De, ``Jammer-assisted resource allocation in
secure {OFDMA} with untrusted users,'' \emph{IEEE Trans. Inf. Forensics Security,} vol. PP, no. 99, pp. 1--1, 2016.

\bibitem{komlosi1993first}
S.~Komlosi, ``First and second order characterizations of pseudolinear
  functions,'' \emph{Europ. J. Operational Res.}, vol.~67, no.~2, pp.
  278--286, 1993.

\bibitem{Bazaraa}
M.~S. Bazaraa, H.~D. Sherali, and C.~M. Shetty, \emph{Nonlinear Programming:
  Theory and Applications}.\hskip 1em plus 0.5em minus 0.4em\relax New York:
  John Wiley and Sons, 2006.

\bibitem{sboyd_subgradient}
S.~Boyd, L.~Xiao, and A.~Mutapcic, \emph{Subgradient methods}, ser. Lecture
  Notes.\hskip 1em plus 0.5em minus 0.4em\relax Stanford Univ., Apr. 2003.

\end{thebibliography}
\end{document}